# Monitoring the conformational dynamics of a single potassium transporter by ALEX-FRET

N. Zarrabi[a], T. Heitkamp[b], J.-C. Greie[b], M. Börsch*[a]

[a]3rd Institute of Physics, University of Stuttgart, Pfaffenwaldring 57, 70550 Stuttgart;
[b]Department of Biology/Chemistry, University of Osnabrück, Barbarastraße 11, 49076 Osnabrück; Germany

**ABSTRACT**

Conformational changes of single proteins are monitored in real time by Förster-type resonance energy transfer, FRET. Two different fluorophores have to be attached to those protein domains, which move during function. The distance between the fluorophores is measured by relative fluorescence intensity changes of FRET donor and acceptor fluorophore, or by fluorescence lifetime changes of the FRET donor. The fluorescence spectrum of a single FRET donor fluorophore is influenced by local protein environment dynamics causing apparent fluorescence intensity changes on the FRET donor and acceptor detector channels. To discriminate between those spectral fluctuations and distance-dependent FRET, alternating pulsed excitation schemes (ALEX) have recently been introduced which simultaneously probe the existence of a FRET acceptor fluorophore. Here we employ single-molecule FRET measurements to a membrane protein. The membrane-embedded KdpFABC complex transports potassium ions across a lipid bilayer using ATP hydrolysis. Our study aims at the observation of conformational fluctuations within a single P-type ATPase functionally reconstituted into liposomes by single-molecule FRET and analysis by Hidden-Markov-Models.

## 1. INTRODUCTION

Ions are important for living cells as, for example, enzyme cofactors and co-substrates for sym- and antiport processes, and they are involved in the homeostasis of intracellular pH. To maintain the concentration of ions against a concentration difference across the membranes, phosphorylated-type (P-type) ATPases are involved in pumping cations [1]. For example, the electrical signals in brains and hearts are generated by sodium-potassium pumps, and muscle contraction requires calcium ions which are pumped by SERCA [2-4]. These nanomachines use the energy provided by the hydrolysis of adenosine triphosphate, ATP, for the transport.

In *Escherichia coli* the membrane-embedded KdpFABC complex uses ATP hydrolysis to transport potassium ions across the plasma membrane [5-7]. This complex belongs to the group of P-Type ATPases and consists of four subunits (Fig. 1a). Subunit KdpB has a molecular weight of 72 kDa and contains the ATP binding site, whereas subunit KdpA is the ion-translocating subunit. The structural features and conformational dynamics of KdpB can be anticipated using a SERCA-derived homology model (Fig. 1b). Three domains are found in the extra-membranous part of KdpB, the nucleotide binding domain N, the phosphorylation domain P, and the actuator domain A.

Characteristic for the reaction cycle of all P-type ATPases is the presence of two major conformational states named E1 and E2. A presumed E1 conformation of KdpB is shown in Figure 1b. Here, the N domain does not contain a nucleotide (ATP or ADP). The N domain and the A domain are far apart from each other. Upon binding of ATP to the binding site at the N domain, the N domain approaches the P domain (by moving to the right in Figure 1b). ATP gets hydrolyzed and the phosphate group of ATP is transferred to the P domain. Simultaneously, the actuator domain A rotates. The new conformation is called E1~P.

..............................................................................................................................................................

*m.boersch@physik.uni-stuttgart.de; phone (49) 711 6856 4632; fax (49) 711 6856 5281; www.m-boersch.org

In the next step the ADP is released from the nucleotide binding site and the K$^+$ ion is transported in the KdpA subunit. Therefore, the N domain in the E2~P conformation has to move backwards, i.e. away from the A domain. Now, the phosphate group is released and the new configuration called E2 has an empty nucleotide binding site in KdpB as well as an empty potassium binding site in KdpA.

As anticipated from the proposed SERCA reaction cycle, the E2 state changes to the E1 state as the final step in the reaction cycle upon K$^+$ binding to KdpA. The major difference between the E1 and the E2 conformations is not only the orientation of the actuator domain A, which switches back, but also a concerted tilting motion of all three domains of KdpB. However, it is not known how flexible these individual conformational states are.

To observe the conformational dynamics of KdpB we genetically introduced two accessible cysteines for stochastic labeling with the two fluorophores Alexa488-maleimide and Atto655-maleimide. One fluorophore was attached to the N domain at position 407 and the other in the A domain at position 150 (Fig. 1b). Single KdpFABC complexes were reconstituted into lipid vesicles (liposomes). The KdpFABC containing liposomes were diluted in buffer to pass the confocal detection volume one after another, and Förster-type resonance energy transfer, FRET, between the two dyes on KdpB was probed by pulsed alternating laser excitation (ALEX [8] or PIE [9], respectively). Fluorescence lifetime information from the FRET donor Alexa488 was used to estimate the minimal number of distinguishable conformational states. For detecting fast fluctuations in the millisecond time range Hidden-Markov-Models were evaluated and the dwell time resolution for conformational states was scrutinized by simulated FRET time trajectories.

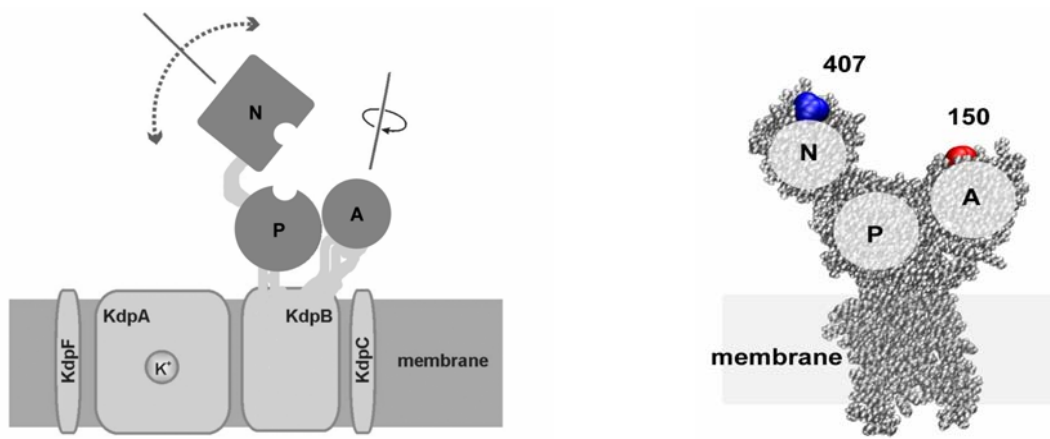

Fig. 1. Left, the potassium transporter KdpFABC from *Escherichia coli* consists of four membrane-bound subunits KdpF, KdpA, KdpB and KdpC. Potassium ions are translocated by subunit KdpA across the lipid membrane. Subunit KdpB is composed of a membrane-embedded protein part, the nucleotide binding domain **N**, the phosphorylation domain **P**, and the actuator domain **A**. The putative domain motions during the reaction cycle of subunit KdpB are indicated by dotted and solid arrows. During K$^+$ transport the N domain moves towards the A domain, and back (dotted arrows). In addition, the A domain is expected to rotate between two orientations relative to the P domain. Right, KdpB homology model in the E1 conformation (as derived from the Ca$^{2+}$ transporter SERCA) showing the positions of the two exposed cysteines at amino acid positions 150 and 407. These cysteines were stochastically labeled with either Alexa488-maleimide or Atto655-maleimide, respectively, for the intramolecular single-molecule FRET measurements.

## 2. MATERIALS AND METHODS

### 2.1. FRET-labeled subunit KdpB in liposomes

The biochemical methods to obtain the cysteine mutants of KdpB for the FRET labeling are to be published elsewhere. KdpFABC complexes containing subunit KdpB with two exposed cysteines were purified from *Escherichia coli* membranes according to published procedures and reconstituted to preformed lipid vesicles (Avanti lipids). Stochastic labeling of the two cysteines was achieved using the liposome-embedded protein. In the first step, sub-stoichiometric amounts of Alexa488-maleimide (Molecular Probes) were added to the proteoliposomes and reacted. Afterwards, Atto655-maleimide (Atto-tec) was added and reacted with the remaining cysteines, i.e., at the second residue position. Excess of unbound fluorophores was removed by subsequent passages through Sephadex G 50 columns. Fluorescent impurities in the TRIS buffer solutions were removed by active charcoal (Merck) treatment for several hours. The FRET-labeled KdpB in liposomes were diluted to final concentrations of about 100 pM immediately before the single-molecule FRET measurements.

## 2.2. Single-molecule FRET measurements using pulsed ALEX with 488 nm and 635 nm

Single-molecule FRET measurements were accomplished on a custom-designed confocal microscope based on an inverted Olympus IX 71 [10-16]. Fiber coupled ps-pulsed laser sources were available at 488 nm (PicoTa 490, operated at 40 MHz repetition rate; Picoquant, Berlin, Germany) and 635 nm (LDH-P-C-635B, 40 MHz repetition rate; Picoquant), which were triggered by the laser driver unit PDL-808 SEPIA (2-channel version, Picoquant). For alternating laser excitation the red laser pulse was delayed by 15 ns with respect to the 488 nm pulse using a prolonged trigger cable.

Laser beams at 488 nm and 635 nm were compressed to beam diameters of about 3 mm for the solution measurements, and overlaid by reflection from (or transmission through) a dichroic beam splitter (DXCR 540, AHF, Tübingen, Germany), respectively. In epi-fluorescence configuration, both beams were re-directed by a dual band dichroic filter (HC dual line beam splitter 488/633-638, AHF) and focused in the buffer droplet by a water immersion objective (UPlanSApo 60xW, 1.2 N.A., Olympus, Hamburg, Germany). Fluorescence of FRET donor and acceptor dyes were detected by two avalanche photodiodes simultaneously (AQR-14, Perkin Elmer, Germany). After passing the 150 μm pinhole, fluorescence was separated by a dichroic HQ640LP (AHF) and was detected in the FRET donor channel from 497 nm to 567 nm (interference filter HQ 532/70, AHF) and in the acceptor channel above 665 nm (long pass filter HQ665LP, AHF). Single photons were registered by a time correlated single-photon counting PC card (SPC-630, Becker&Hickl, Berlin, Germany) combined with an 8-channel router (HRT-82, Becker&Hickl). For triggering the SPC-630 card, the trigger output of the PDL-808 SEPIA was used in combination with the LTT100 sync adaptor (Picoquant).

Photons were stored in the FIFO mode of the TCSPC card. Therefore, the so-called 'macro time' of each photon had a time resolution of 50 ns, which was used for 1 ms binning of the fluorescence intensity data and for fluorescence correlation spectroscopy. The 'micro time' of each photon (arrival time with respect to the laser pulses) was used to calculate the fluorescence lifetime histograms within each selected photon burst or time bin. Furthermore, the 'micro time' information of the photons in the FRET acceptor channel was applied to discriminate between photons detected as a result of FRET following excitation at 488 nm, and photons by direct excitation of the FRET acceptor with the red laser pulse.

Time trajectories of the FRET signals as well as the FRET acceptor control were analyzed by the custom-designed software 'Burst_Analyzer' [12-16]. At first, photon bursts of FRET-labeled KdpB in liposomes were selected by fluorescence intensity thresholds using the acceptor control time trajectory (Fig. 2). That is, each burst had to exceed 20 counts per ms for direct excitation of the acceptor and on average 5 kHz (i.e., 5000 counts per second or 5 counts per millisecond, respectively) throughout the burst. Thereby, photon bursts showing apparent FRET caused by spectral fluctuations of the FRET donor, but without a FRET acceptor dye bound to the protein, could be excluded from further analysis (Fig. 2b). In addition, photon burst showing fluorescence from the FRET acceptor after direct excitation at 635 nm, but showing no fluorescence following the excitation pulse at 488 nm, could be removed by a minimum threshold in the FRET donor channel of 5 kHz on average (Fig. 2c).

## 2.3. Monte Carlo simulation of single-molecule FRET data and Hidden Markov Analysis

To check the time resolution limits of detecting a putative conformational state by intensity-based FRET, Monte Carlo simulations of a virtual FRET labeled particle were computed as described previously [15]. Briefly, a single particle with a diffusion constant according to the size of the liposomes moved randomly through a box that included a three-dimensional Gaussian-shaped excitation / detection volume. The detection volume was sized according to the experiment. For the particle, FRET states were set with transition rates for sequential FRET changes. A Poisson-distributed background signal was added to the simulated fluorescence intensities to mimic the experimental conditions. Photon bursts were recovered from the resulting 2-channel time trajectories by intensity threshold criteria.

The FRET fluctuations of the experimental and the simulated FRET data within a burst were analyzed using a Hidden Markov Model approach [17-22] with time-binned intensity data. The three-state model had been used previously to analyze single-molecule FRET data of the rotary motions of subunits in $F_oF_1$-ATP synthase [16]. For this work, the HMM has been refined to overcome the problems of FRET state turning points within one time bin. These time bins occur very often and might result in an assignment of a hidden state to an averaged observable. However, state transitions are indicated by a reduced corresponding likelihood value of the Markov chain on those time bins. Previously, we used the summed number of FRET donor and acceptor photons per time bin as the weighting information for the accuracy of the measured proximity factor. Including the reduced likelihood value at the transition points as weighting information diminished the contribution of time bins with a high probability of an inherent state transition.

## 3. RESULTS

We aimed at monitoring conformational dynamics related to the reaction cycle of the two extra-membranous domains N and A of the potassium transporter subunit KdpB from *Escherichia coli*. Therefore, two exposed and reactive cysteines at residue position 407 and 150, respectively, were introduced genetically. To find an optimal FRET donor we evaluated the photophysical and labeling properties of three fluorophores suitable for single-molecule spectroscopy: Alexa488, Bodipy-FL, and Atto-488. Large spectral fluctuations were observed after binding to KdpB for the two fluorophores Bodipy-FL and Atto-488. In addition, the labeling efficiencies were found to be rather low, and non-specific binding to the lipid membrane complicated the removal of excess dye. Therefore, Alexa488 was chosen as the FRET donor. Atto655 was employed as the FRET acceptor because of its known photostability compared to Cy5, being hydrophilic and exhibiting an acceptable fluorescence quantum yield of about 30 % with a fluorescence lifetime of 1.9 ns according to the supplier. The estimated Förster radius for this FRET pair is $R_0=5.0$ nm. Our fluorophore labeling strategy resulted in a stochastic labeling of the KdpB with either only one dye or two dyes attached, while mixed labeling yielded the FRET-labeled KdpB, but double-labeling with either FRET donor or FRET acceptor dye, respectively, resulted in a non-FRET labeled KdpB.

### 3.1. Time-resolved single-molecule FRET data

To select the FRET-labeled KdpB in liposomes from KdpB only labeled with either Alexa488 or Atto655, the alternating laser excitation approach was used (ALEX or PIE, respectively). Two laser diodes, the amplified frequency-doubled PicoTA 490 at 488 nm and the high power laser diode LDH-635 at 635 nm, were pulsed sequentially with the red laser delayed by 15 ns. Excitation power of the PicoTA 490 was adjusted to 150 μW and the excitation power of the red laser was reduced to 25 μW. To prove the colocalization of the overlaid excitation foci, a sample with immobilized fluorescent microbeads on a glass cover slide was scanned in three dimensions and image stacks from the two lasers were compared.

Subsequently, a mixture of the fluorophores rhodamine 110 and Atto655 was diluted in $H_2O$ to picomolar concentrations for single-molecule measurements. The diffusion time of rhodamine 110 excited at 488 nm was about 170 μs as calculated from the autocorrelation function. The diffusion time of the slightly larger Atto655 dye excited at 635 nm was about 350 μs. The amplitude of the cross correlation function was negligible as expected for two independently diffusing species. The fluorescence lifetime of rhodamine 110 was 4.05 ns using a maximum likelihood estimator [23]. For Atto655, the lifetime was found to be 1.8 ns as reported.

The KdpFABC complexes containing the double-labeled KdpB in liposomes were diluted to about 100 pM with TRIS buffer in the presence of ATP and KCL and measured in 50 μl droplets. As the protein traversed the confocal excitation volumes, a burst of fluorescence photons was created (Fig. 2). The diffusion time of KdpFABC in liposomes was about 30 ms, when calculated from the autocorrelation function of the FRET acceptor (directly excited with 635 nm), indicating liposome diameters between 100 nm and 200 nm. In the fluorescence time trajectories, searching for single FRET-labeled KdpFABC transporters was performed using an intensity threshold of at least 5 KHz on average in the FRET acceptor channel probed by direct excitation.

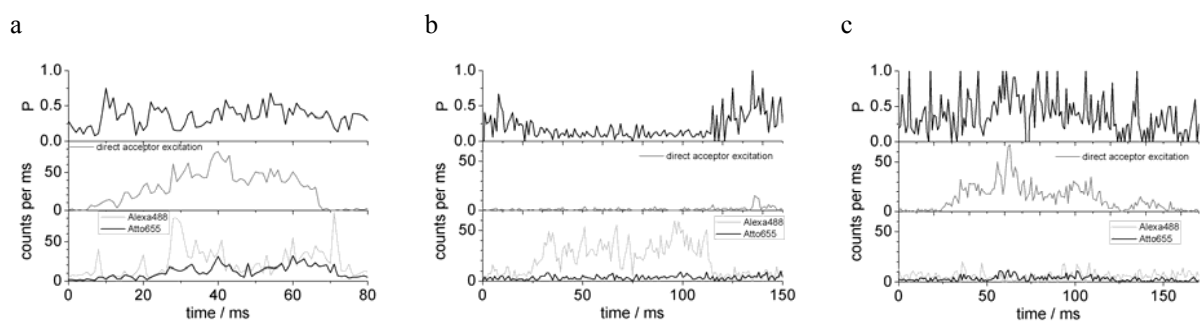

Fig. 2. Single KdpFABC complexes in a liposome labeled with one or two fluorophores in the KdpB subunit, respectively, in the presence of ATP and KCL. Lower traces show the fluorescence intensities of FRET donor (light grey) and FRET acceptor (black). The fluorescence intensity of the FRET acceptor by direct excitation is shown above. The proximity factor P is calculated from the FRET trajectories and depicted in the upper traces. **a**, single FRET-labeled KdpFABC showing fluctuations in the proximity factor; **b**, single FRET donor-only labeled KdpFABC without an acceptor fluorophore; **c**, single FRET acceptor-only labeled KdpFABC without the FRET donor. Time binning was 1 ms.

Donor-only labeled KdpFABC (Figure 2b) was omitted from further analysis. Acceptor-only labeled KdpFABC was removed by an intensity threshold of at least 5 KHz on average in the FRET donor channel following the laser pulse at 488 nm (Figure 2c). The data set consisted of 470 photon bursts for FRET analysis.

Within these photon bursts, the fluorescence intensities of FRET donor and acceptor fluctuated (Fig. 2a). As a measure of the FRET efficiency we calculated a proximity factor P

$$P = \frac{I^*_A}{I^*_A + I^*_D} \quad (1)$$

with $I^*_A$ and $I^*_D$, non-corrected fluorescence intensities in the FRET acceptor or FRET donor channel, respectively. The proximity factor is related to the FRET efficiency, $E_{FRET}$

$$E_{FRET} = \frac{I_A}{I_A + \gamma I_D} \quad (2)$$

with $I_A$ and $I_D$, background corrected fluorescence intensities in the FRET acceptor or FRET donor channel, and $\gamma$ being a correction factor, which consists of the quantum yields of the dyes and the detection efficiencies of both detection channels. Alternatively, the lifetime information from the FRET donor dye can be used. Increasing FRET efficiency is reflected by a reduction of the FRET donor lifetime

$$E_{FRET} = 1 - \frac{\tau_{DA}}{\tau_{D_0}} \quad (3)$$

with $\tau_{DA}$, the fluorescence lifetime of the FRET donor in presence of the acceptor, $\tau_{D_0}$ is the fluorescence lifetime of the donor in the absence of a FRET acceptor or other quenchers in the local environment.

In order to prove FRET being the cause of the fluctuations in the fluorescence time trajectories of KdpB, the FRET donor lifetimes were calculated for different proximity factors within the selected photon bursts. FRET donor photons were assigned to 10 intervals of P (0 to 0.1, 0.1 to 0.2, ..., and 0.9 to 1) as shown in Fig. 3b and added into the histogram. For comparison, the lifetimes of FRET donor-only labeled KdpFABC in liposomes were measured from photon bursts. FRET donor fluorophores attached to residue position 150 in KdpB showed the identical fluorescence decay curves as those attached to position 407 (Fig. 3a). For both positions, the bi-exponential fittings with decay times of 0.7 ns and 3.8 ns were superior to a monoexponential fit. The mean lifetime for Alexa488 bound to KdpB was 3.0 ns. The existence of a short lifetime component with an amplitude of 25 to 30 percent could indicate a reversible quenching by the local protein environments.

Fluorescence lifetime decays were fitted separately depending on the P intervals for the FRET-labeled KdpB. Monoexponential fittings were sufficient for 0 < P < 0.1 resulting in $\tau$ = 3.47 ns and for 0.1 < P < 0.2 yielding $\tau$ = 3.15 ns. For P > 0.2, all decay curves had to be fitted bi-exponentially with $\tau_1$ = (3.6±0.2) ns and $\tau_2$ = (0.57±0.1) ns, but amplitudes increased for the short lifetime component (Fig. 3c). The fluorescence lifetime of the FRET acceptor bound to KdpB, directly excited at 635 nm, was $\tau$ = (2.2±0.1) ns for all proximity factor intervals.

The broad distribution of proximity factors had a maximum for 0.2 < P < 0.3 (Fig. 3d). Adding up the FRET donor and acceptor photons for each P interval resulted in a constant level of mean fluorescence intensity for all proximity factor intervals with P > 0.2 (Fig. 3e). However, for P < 0.2 the mean fluorescence intensity was significantly higher indicating some bright fluorescence impurities, which might be due to the presence of single, fast diffusing fluorescent contaminations.

The mean fluorescence lifetimes followed both the relations of FRET efficiency calculated by fluorescence lifetime (equation 3) and by intensity ratio (equation 2). However, the fitting of all decay curves could be interpreted to support a number of three distinguishable lifetimes for all proximity factor intervals, but with varying amplitudes.

To identify possible photophysical quenching of the fluorophores in their local protein environment, the stoichiometry factor S was plotted against the proximity factor. Here, S is defined as

$$S = \frac{I^*_{D,Ex488} + I^*_{A,Ex488}}{I^*_{A,Ex635} + I^*_{A,Ex488} + I^*_{D,Ex488}} \quad (4)$$

with $I^*_{D,ex488}$, non-corrected FRET donor fluorescence after excitation with 488 nm, $I^*_{A,ex488}$, non-corrected FRET acceptor fluorescence after excitation with 488 nm, $I^*_{A,ex635}$, non-corrected FRET acceptor fluorescence after direct excitation with 635 nm. The S-versus-P plot is shown Fig. 4a. All photons of time trajectories were binned to 2 ms intervals and the following thresholds were applied: at least 6 counts and less than 70 counts per ms for $I^*_{D,ex488}$, less than 70 counts per ms for $I^*_{A,ex488}$, and at least 10 counts and less than 70 counts per ms for $I^*_{A,ex635}$, which was multiplied by 1.5. These thresholds exclude the donor-only labeled as well as the acceptor-only labeled KdpB populations from the map. The majority of the remaining FRET-labeled KdpB showed proximity factors between 0.1 and 0.6. However, the proximity factor P exceeded 0.8 for a small fraction of time bins.

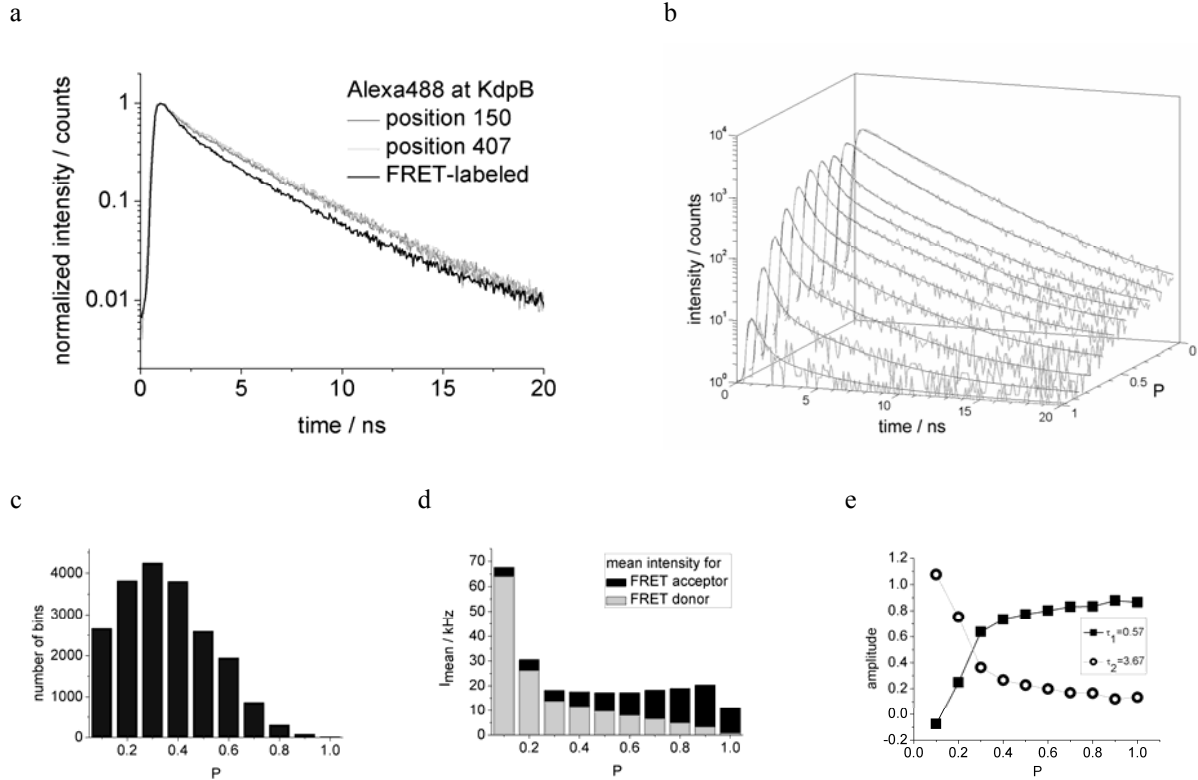

Fig. 3. **a**, fluorescence lifetimes of Alexa488 bound to either position 150 (grey line) or to 407 (light grey line) of KdpB in the absence of a FRET acceptor, or mean lifetimes of all selected photon bursts of FRET-labeled KdpFABC (solid black line). **b**, fluorescence decays and lifetime fits for each of 10 different proximity factor P intervals. P was computed for time bins of 1 ms within selected photon bursts of single FRET-labeled KdpFABC in the presence of ATP and KCl. **c**, distribution of proximity factors in the selected photon bursts. **d**, mean fluorescence intensity summed over all photons in the FRET donor channel (light grey) and acceptor channel (black) for each proximity factor interval. **e**, amplitudes of the short (black squares) and long (open circles) lifetime component in the bi-exponential lifetime fits to the fluorescence decays shown in **b**.

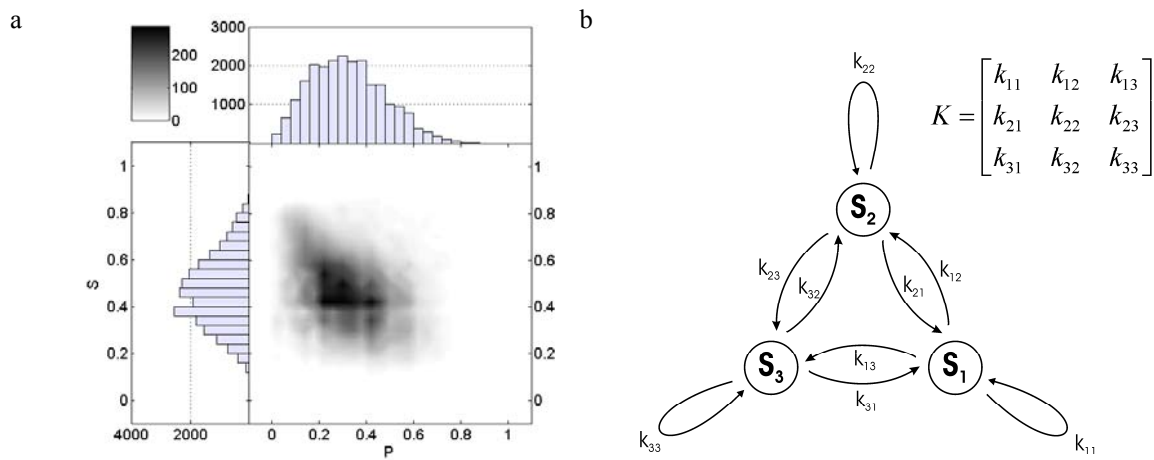

Figure 4. **a**, two-dimensional plot of the stoichiometry factor S *versus* proximity factor P (using all photons in the time trajectories with increased 2 ms time binning and without selecting bursts according to the burst-wise acceptor probing). **b**, three-state Hidden Markov model [15] applied to the FRET level search within the selected FRET-labeled KdpFABC bursts.

### 3.2. Analyzing FRET levels in single bursts by HMM

The fitting results of the FRET donor lifetimes were depending on the proximity factor value of the time bin within the selected photon bursts of KdpFABC. They seemed to support the presence of three FRET states or conformational states, respectively. However, these FRET states could not be assigned manually because the value of the FRET states were unknown as well as the dwell times for each state. The FRET donor lifetime of $\tau = 3.45$ ns corresponded to a proximity factor $P = 0.1$, $\tau = 3.15$ ns could be assigned to $P = 0.2$, and $\tau = 0.6$ ns to $P > 0.8$.

We employed a Hidden Markov model with three hidden states according to Figure 4b. The state S1 was assigned to represent a low FRET efficiency state, S2 was a medium FRET efficiency state, and S3 was a high FRET efficiency state by the initial proximity factor parameters of P=0.2, 0.5 and 0.8. In addition, we tested a cyclic 10-state HMM with equally distributed initial P values. Both HMMs found very similar solutions, that is, fast fluctuations with dwell times about 2 to 15 ms for three distinct hidden states. S1 was attributed to $P = 0.14$ with a dwell time of $t_d = 3.1$ ms (13 % of all assigned states), S2 corresponded to $P = 0.26$ with a dwell time $t_d = 3.9$ ms (38 % of all assigned states), and S3 was found at $P = 0.52$ with a dwell time $t_d = 13$ ms (49 % of all assigned states).

The time binning of the FRET time trajectories has been set to 1 ms, and a Poisson distribution background intensity of 3 kHz was taken into account for the FRET donor channel only. To suppress the influence of the exceptionally brighter time bins associated with proximity factors $P < 0.2$ (see Figure 3d), the selected photon bursts were filtered again using a higher mean intensity of 10 kHz for the directly excited FRET acceptor as well as an additional peak maximum filter of 90 counts per millisecond in the FRET donor channel. Thereby, the distribution of the proximity factors in the S-*versus*-P plot in Figure 4a lost the population with S values about 0.7 and $P \sim 0.1$. By this filtering, 368 photon burst remained out of 470, with 1180 assigned FRET states.

Time constants of the conformational fluctuations, that are shorter than 5 ms or 5 time bins, respectively, are difficult to determine by single-molecule FRET measurements and are arguable [11,12]. Fluorescent impurities with fast diffusion times could cause apparent fluctuations of the FRET efficiencies.

### 3.3. Validating HMM results by simulated single-molecule FRET data

To test the robustness of short dwell times as found by the HMM analysis, we generated artificial FRET data sets mimicking the experimental data. In the Monte Carlo simulations, a single FRET-labeled particle was diffusing through a box of 301 fl with an included 3D Gaussian-shaped 'excitation volume' of 5.84 fl. The particle moved with a diffusion constant $D = 6.0 \cdot 10^{-8}$ cm$^2$/s through the box as expected for a liposome-like particle. Particle detection was possible within the 'excitation volume'. The sum brightness of both fluorophores for FRET was set to 100 kHz in the center of the 'excitation volume' and 2-channel fluorescence time trajectories were recorded. The particle changed its FRET values according to the given three FRET states S1 at $P = 0.15$ (3 ms dwell), S2 at $P = 0.35$ (3 ms dwell) and S3 at $P = 0.55$ (8 ms dwell). A preferred FRET state sequence was applied by S1 → S3 → S2 → S1 → to identify the number of correct assigned states after recovering these states by the HMM. The time resolution in the simulation was set to 100 µs for 10 trajectories with a length of 100 s each. Subsequent re-binning the simulated FRET data allowed for evaluation of the lower time resolution limits of the HMM analysis. A Poisson distributed background signal of 4 kHz was added on both channels.

The simulation generated time trajectories with well-separated single photon bursts showing the characteristic fluctuations of the fluorescence intensities due to the Brownian motion through the excitation volume. Within the bursts, the relative fluorescence intensities changed indicating the FRET efficiency fluctuations of the system. For burst identification, threshold criteria were used with minimum average count rates of 10 kHz for FRET donor as well as the FRET acceptor channel. The three-state HMM with initial proximity factor parameters of P=0.2, 0.5 and 0.8 was applied. The HMM assigned 3568 FRET states in 993 bursts, and obtained S1 with P=0.14 (2.7 ms dwell, 20 % of assigned states), S2 with $P = 0.32$ (3.9 ms dwell, 34% of states) and S3 with P= 0.55 (9.8 ms dwell, 46 % of states). For 73 percent of all time bins, the respective state was assigned correctly by the HMM, and the dwell times were found in good agreement with the simulation settings.

In another simulation we chose longer dwells of 10 ms for $P = 0.2$ and $P = 0.5$ and a very short 1 ms dwell for $P = 0.8$. This simulation was intended to represent a FRET state sequence including a transient intermediate state with a dwell time equal to a single time bin of the measurement. A preferred state sequence was given again by S1 → S3 → S2 → S1 → . To identify a burst, minimum average count rates of 10 kHz for FRET donor as well as the FRET acceptor channel were required. The retrieval probabilities for the hidden states were very good, with correct assignment to states for more than 85 % (1924 states in 698 bursts). The HMM recovered P = 0.198

(11.3 ms dwell, 42 % of assigned states), P=0.495 (12.6 ms dwell, 53 % of assigned states) and in particular the short state P3 = 0.775 (0.91 ms dwell, 5 % of assigned states).

## 4. DISCUSSION

Based on sequence comparison and biochemical analyses, the membrane-embedded KdpFABC complex belongs to the group of P-type ATPases. In contrast to other well-characterized members of this family (for example the sarcoplasmic reticulum $Ca^{2+}$ ATPase, SERCA), KdpFABC consists of four subunits, in two of which the sites of substrate transport (KdpA) and ATP hydrolysis (KdpB) are spatially separated. Subunit KdpB comprises all conserved sequence motifs of P-type ATPases, whereas the sequence of KdpA strongly mimics that of MPM-type $K^+$ channels. Like all other P-type ATPases, the KdpFABC complex is expected to undergo a reaction cycle during catalysis resulting in conformational dynamics.

We apply a single-molecule FRET approach to get a first clue to the conformational dynamics of KdpB in real time within a single transporter. Using pulsed laser excitation, the fluorescence lifetime information of the FRET donor fluorophore was accessible and could be compared to the ratiometric fluorescence intensity based FRET efficiency calculations of the same photon burst. Here, pulsed laser excitation of the FRET donor was combined with direct excitation of the FRET acceptor using a second, delayed laser pulse. This ALEX approach allowed discriminating apparent FRET efficiencies solely caused by spectral fluctuations of the FRET donor fluorophore.

In addition, direct probing of the FRET acceptor by ALEX could be used to identify photon bursts of KdpFABC molecules traversing the center of the confocal laser excitation volume. The FRET trajectories of those bursts contain the most valuable FRET information even if the brightness is not as high. Especially, photon burst exhibiting medium to high FRET efficiencies and, correspondingly, lower count rates from the FRET donor, could be classified according to the FRET information quality. Increasing the thresholds for the directly-excited FRET acceptor count rates slightly reduced the overall number of FRET-labeled photon bursts but shaped the proximity factor distribution of the remaining photon burst. Most importantly, the fraction of FRET bursts with $P < 0.2$ but significantly higher mean values of the sum brightness of FRET donor an acceptor channel (shown in Fig. 3d) was identified to exhibit low FRET acceptor count rates. By applying an additional peak maximum threshold of 90 kHz for the FRET donor channel. These photon burst could originate from liposomes containing more than one KdpFABC or additional fluorophores emitting in the FRET donor channel. These bursts were removable from further analysis.

As the first step to identify the number of distinguishable FRET states in KdpFABC during ATP hydrolysis, we analyzed the FRET donor lifetimes in photon bursts with an assured presence of the FRET acceptor. Each time bin within a selected photon burst was classified according to the proximity factor value $P = I^*_A/(I^*_D+I^*_A)$ without corrections for the fluorescence background signals. FRET donor photons were summed up for each of the 10 different intervals of P. Fluorescence decays were well fitted using one or two exponential decay components. For $P > 0.3$, all decay curves contained a shorted-lived and a long-lived component with 0.6 ns and 3.6 ns but varying amplitudes. The average brightness, calculated from the sum of FRET donor and FRET acceptor photon divided by the length of the burst, was found to be constant up to $P = 0.9$. Accordingly it can be concluded, that fluctuations between two different FRET efficiency states, which are faster than the time bin, give rise to an averaged FRET efficiency. The precision of the lifetime information is in the picoseconds to nanoseconds range. Therefore, lifetime changes can identify conformational fluctuations in the sub-millisecond range causing FRET efficiency changes. However, a large number of photons is required for a high reliability of mono- or bi-exponential decays, and, therefore, addition of many time bins was necessary. In contrast, ratiometric fluorescence intensity-based information is able to identify conformational changes of protein states in the millisecond time range without the need of sampling. To overcome the time resolution limitations of the ratiometric FRET data analysis, a photon-by-photon analysis of FRET efficiency changes has been suggested.

In this work, a search algorithm based on a three-state Hidden Markov model was employed to find conformational states of KdpFABC. The associated distinct FRET efficiencies were calculated from fluorescence intensities. Following our previous attempts to analyze subunit rotation in FRET-labeled $F_oF_1$-ATP synthase, the HMM approach was refined by including the reduced likelihood value of the Markov chain at the transition points of two states. Thereby, the averaged proximity factor values of these time bins were omitted from state assignments by the HMM.

With this modified HMM, we found three possible FRET states for the double-labeled KdpFABC during ATP hydrolysis. The FRET states were short-lived (between 3 and 13 ms) and associated with proximity factors $P = 0.14$, 0.26, and 0.52. Applying HMMs with more hidden states did not change the result of three dominating states plus negligible other states with lower frequencies. However, also a two-state HMM was evaluated yielding FRET states with $P = 0.17$ (3 ms dwell, 39 % of assigned states) and with $P = 0.47$ (13 ms dwell, 61 % of assigned states). As some of the assigned FRET states were short-lived in comparison to the 1 ms time bin of the FRET data, the time resolution limits of HMM results were checked by artificial FRET data. Monte Carlo simulations provided data of a virtual 3-state system with distinct FRET efficiencies and short dwell times. HMM recovery showed that short-lived states can be identified. As the next step in gaining evidence for fast submillisecond fluctuations of KdpFABC, we will analyze the cross correlation functions of the fluorescence trajectories of FRET donor and acceptor within selected photon bursts.

Some complications in the real FRET data could not yet be compensated by ALEX and FRET acceptor probing. Mislabeling of native cysteines in the KdpB protein, fluorescence impurities in the lipid bilayer or inside the liposome or the concomitant presence of two reconstituted KdpFABC complexes in a single liposome will result in proximity factors which are not related to the FRET-dependent values. During HMM analysis, these photon burst will contribute to the overall state assignment and are weighted by their brightness.

To attribute the putative conformational fluctuations of the N domain and the A domain of KdpB, trapping of individual catalytic states with known conformations of the soluble domains of KdpB is required. Conformational dynamics can be restricted by either specific inhibitors or, more generally, by cooling. Surface immobilization could allow for increased observation times of KdpB during turnover. Furthermore, the use of other FRET donor fluorophores instead of Alexa488 could help to discriminate between photophysical artifacts and conformational changes of the two domains of the subunit. Based on our results abtained so far, this might pave the road to monitor conformational dynamics in real time within a single functionally assembled P-type ATPase.

## 5. REFERENCES

1. J.V. Møller, B. Juul, M. le Maire, M., "Structural organization, ion transport, and energy transduction of P-type ATPases", Biochim. Biophys. Acta 1286, 1-51 (1996).
2. C. Toyoshima, M. Nakasako, H. Nomura, H. Ogawa, H., "Crystal structure of the calcium pump of sarcoplasmic reticulum at 2.6 Å resolution", Nature 405, 647-655 (2000).
3. C. Toyoshima, T. Mizutani, "Crystal structure of the calcium pump with bound ATP analogue", Nature 430, 529-535 (2004)
4. C. Olesen, M. Picard, A.M. Winther, C. Gyrup, J.P. Morth, C. Oxvig, J.V. Møller, P. Nissen, "The structural basis of calcium transport by the calcium pump", Nature 450, 1036-1044 (2007).
5. M. Gaßel, A. Siebers, W. Epstein, K. Altendorf, "Assembly of the Kdp complex, the multi-subunit $K^+$-transport ATPase of Escherichia coli", Biochim. Biophys. Acta 1415, 77-84 (1998).
6. M. Haupt, M. Bramkamp, M. Coles, K. Altendorf, H. Kessler, "Inter-domain motions of the N-domain of the KdpFABC complex, a P-type ATPase, are not driven by ATP-induced conformational changes", J. Mol. Biol. 342, 1547-1558 (2004).
7. J.-C. Greie, K. Altendorf, "The $K^+$-translocating KdpFABC complex from Escherichia: a P-type ATPase with unique features", J Bioenerg. Biomembr., DOI 10.1007/s10863-007-9111-0
8. A. N. Kapanidis, N. K. Lee, T. A. Laurence, S. Doose, E. Margeat, S. Weiss "Fluorescence-aided molecule sorting: analysis of structure and interactions by alternating-laser excitation of single molecules", Proc. Natl. Acad. Sci. USA. 101, 8936–8941 (2004).
9. B. K. Müller, E. Zaychikov, C. Bräuchle, D. C. Lamb, "Pulsed interleaved excitation", Biophys. J. 89, 3508-3522 (2005).
10. M. Börsch, M. Diez, B. Zimmermann, R. Reuter, P. Gräber, "Stepwise rotation of the γ-subunit of $EF_oF_1$-ATP synthase observed by intramolecular single-molecule fluorescence resonance energy transfer", FEBS Letters 527, 147-152 (2002).
11. M. Diez, B. Zimmermann, M. Börsch, M. König, E. Schweinberger, S. Steigmiller, R. Reuter, S. Felekyan, V. Kudryavtsev, C. A. M. Seidel, P. Gräber, "Proton-powered subunit rotation in single membrane-bound $F_oF_1$-ATP synthase", Nat Struct Mol Biol 11, 135-41 (2004).
12. B. Zimmermann, M. Diez, N. Zarrabi, P. Gräber, M. Börsch, "Movements of the ε-subunit during catalysis and activation in single membrane-bound $H^+$-ATP synthase", Embo J 24, 2053-63 (2005).
13. M. G. Düser, N. Zarrabi, Y. Bi, B. Zimmermann, S. D. Dunn, M. Börsch, "3D-localization of the a-subunit of $F_oF_1$-ATP synthase by time resolved single-molecule FRET", Progress in Biomedical Optics and Imaging - Proceedings of SPIE 6092, 60920H (2006).
14. N. Zarrabi, B. Zimmermann, M. Diez, P. Gräber, J. Wrachtrup, M. Börsch, "Asymmetry of rotational catalysis of single membrane-bound $F_oF_1$-ATP synthase", Progress in Biomedical Optics and Imaging - Proceedings of SPIE 5699, 175-188 (2005).
15. N. Zarrabi, M. G. Düser, R. Reuter, S. D. Dunn, J. Wrachtrup, M. Börsch, "Detecting substeps in the rotary motors of $F_oF_1$-ATP synthase by time resolved single-molecule FRET", Progress in Biomedical Optics and Imaging - Proceedings of SPIE 6444, 64440E (2007).

16. N. Zarrabi, M. G. Düser, S. Ernst, R. Reuter, G. D. Glick, S. D. Dunn, J. Wrachtrup, M. Börsch, "Monitoring the rotary motors of single $F_oF_1$-ATP synthase by synchronized multi channel TCSPC", Proceedings of SPIE 6771, 67710F (2007).
17. K. P. Murphy, Computing Science and Statistics 33, (2001).
18. S. A. McKinney, C. Joo, T. Ha, "Analysis of single-molecule FRET trajectories using hidden Markov modeling", Biophys. J. 91, 1941-1951 (2006).
19. A. J. Viterbi, "Error Bounds for Convolutional Codes and an Asymptotically Optimum Decoding Algorithm", Ieee Transactions on Information Theory It13, 260-269 (1967).
20. L. E. Baum, T. Petrie, G. Soules, N. Weiss, "A Maximization Technique Occurring in Statistical Analysis of Probabilistic Functions of Markov Chains", Annals of Mathematical Statistics 41, 164-171 (1970).
21. R. T. T. Hastie, and J. Friedman, "The Elements of Statistical Learning: Data Mining, Inference, and Prediction", Springer-Verlag, New York, 2001.
22. D. Talaga "Markov processes in single molecule fluorescence", Curr Opin Colloid In 12, 285-296 (2007).
23. J. Enderlein, R. Erdmann, "Fast fitting of multi-exponential decay curves", Optics Comm., **134**, 371-378, (1997).